# 'Genesis': a takeover from field-responsive matter?

Gargi Mitra-Delmotte[1], Ph.D., and Asoke N. Mitra[2]

[1]39 Cite de l'Ocean, Montgaillard, St.Denis 97400, REUNION;
e.mail: gargijj@orange.fr ;

[2]Emeritus, Department of Physics, Delhi University; 244 Tagore Park, Delhi -110009. INDIA;
e.mail: ganmitra@nde.vsnl.net.in

**Abstract:** Cairns-Smith (2008) has argued for a pre-Darwinian era, with a simpler basis for life's functioning via primitive "crystal genes" (information transfer, kinetic control on metabolic reactions). At the other extreme, guided by the structural similarity of clusters in early-evolved enzymes to iron-sulphide minerals like greigite, the hydrothermal mound scenario of Russell and coworkers (1994) presents how non-equilibrium forces rooted in geochemistry could be extrapolated to understand the metabolic functioning of living systems. The informational vs metabolic aspects of life in these respective scenarios can be linked together via a framboid-based theory of Sawlowicz (2000), as these assemblies typically form in colloidal environments. In this background, we consider the ramifications of a magnetic rock field on the mound scenario, asking if soft matter assemblies are compatible with a coherent order.

Recognizing that "Darwinian evolution is a kind of natural engineer that has given rise to the seemingly purpose-built features" of living systems, Cairns-Smith (2008) has elegantly framed the question of life's emergence as, "How might a Darwinian evolution have started most easily, based on whatever substances were appropriate and available?" But he disagrees with the frequent assumption "that the kind of materials on which evolution depends today would have been involved from the outset". He proposes instead a pre-Darwinian era "during which the essentials of our system were invented, and through which our current "molecules of life" acquired their significance". In particular, information patterns in "crystal genes" are imprinted in the arrangement of consecutive layers, differing in features like chemistry/orientation. Here, instead of "sugar-phosphate strings" preserving a DNA sequence, unit layers of similar thickness are linked by ordinary interlayer forces. Although less efficient than DNA as an information store, a stack of ~100 units of two different kinds has as many as $10^{30}$ permutations! An important aspect of the "primitive-crystal–gene concept" is that organic metabolism was not needed at the very outset of evolution. A primitive sort of metabolism ("chemical reactions under genetic control") is thought to have been initiated in catalytic edge sites in the primitive genetic material; and organics in the local surroundings that were modified by these reactions assisted the propagation of the latter. Further, lattice-vibrations in crystals could have provided enzyme-like action for taking reactants over kinetic barriers en route to the RNA world. But, although the latter helps simplify the paradox of which process-- life's Darwinian evolution or its complex machinery—may have preceded the other, it requires a geochemically implausible supply of RNA monomers (Cairns-Smith 2008).

In that case, life being a history-dependent phenomenon, traces of any inorganic matter would provide clues into the structure of minerals that could have been involved in life's creation. Indeed, iron sulphur clusters are seen ubiquitously across kingdoms of living systems today and carry out a variety of roles, such as electron transfer, radical generation, sulfur donation, control of protein conformational changes associated with signal transduction, to name some (Allen 1993; Johnson 1996; Beinert et al 1997; Baymann et al 2003; Dos Santos and Dean 2008). And, what's more, the hydrothermal mound scenario, envisaged on the floor of the Hadean Ocean, considers clusters in early evolved enzymes in terms of their structural similarity to iron-sulphide minerals that along with some other transition metal compounds could have been geo-chemically available for carrying out similar catalytic type of functions, and acted as proto-metabolically relevant catalysts (Russell and Martin 2004; Martin and Russell 2007; McGlynn et al 2009; Nitschke and Russell 2009). Here, the confrontation of geo-fluids at different pH across a colloidal FeS membrane provides insights into the essential dependence of biological processes on a proton-motive force (chemiosmotic gradient) across cell membranes, which is a more or less invariant mechanism of energy conversion across living forms (Russell and Hall 1997). This picture also gives insights from an evolutionary perspective into how simpler metabolic cycles led to more complex ones (Lane et al 2010). Furthermore, chemicals involved in replication, also started appearing and mutually helped each other's evolution (Milner-White and Russell 2005; 2010).

In this manner, soft reproducing membranous compartments capable of concentrating a rich variety of abiogenics, and with non-equilibrium forces rooted in geochemistry, could address the origins of metabolic functioning of living systems. On the other hand, we seem to be losing sight of the "genetic material" (see above) where much of the concept of control—whether information transfer from patterns imprinted in crystal layers, or the kinetic control on metabolic reactions via lattice vibrations—ideally depends on the *coherent* constituent organization of crystals.

The informational aspects of 'crystal-templates' and metabolic issues addressed in the mound scenario can interestingly be treated together, as suggested by Sawlowicz (2000) in his framboid-based theory. Now, the mound scenario shows access to at least two kinds of inorganic mineral networks: (i) membranes which are understood to age and rigidify into chimneys, and (ii) pyrite framboids (see Figure1; Boyce et al. 1983; Boyce 1990; Larter et al. 1981). And earlier workers looking for clues for stability in natural assemblies have taken keen interest in framboids sprinkled across the cosmos. The assembly of these forms, displayed by structurally different minerals, takes place via interplay of *physical* forces, where attractive ones aiding surface minimization of colloidal-particle assemblies could be provided by surface-tension, magnetic-dipolar forces, etc. Sawlowicz's (2000) recent theory of the origins of life is based on the bio-potential of constituent microcrystal surfaces, presence of catalytic metals, fractal structures, to name some of the fascinating features of framboids. In particular, he emphasizes the aspect of *scale-free* ordering that has been observed in some of these assemblies, consistently with the nested order of living systems (Harold 2005). These natural assemblies show the possibility of access to a *stable fractal organization* by dissipating energy. (This can be contrasted to other natural non-equilibrium systems where either the ordering is not of a heirarchial nature, or fractal patterns are not stable, whereas the more familiar heirarchial organization of domains in a magnet is only an example of an equilibrium, albeit symmetry-broken, system). It may be noted in this context that the past two decades have seen much interest in colloid-aggregation –a far-from-equilibrium kinetic process, and in particular those in the presence of long range dipolar forces (Pastor-Satorras and Rubí 1998; Martínez-Pedrero 2008)

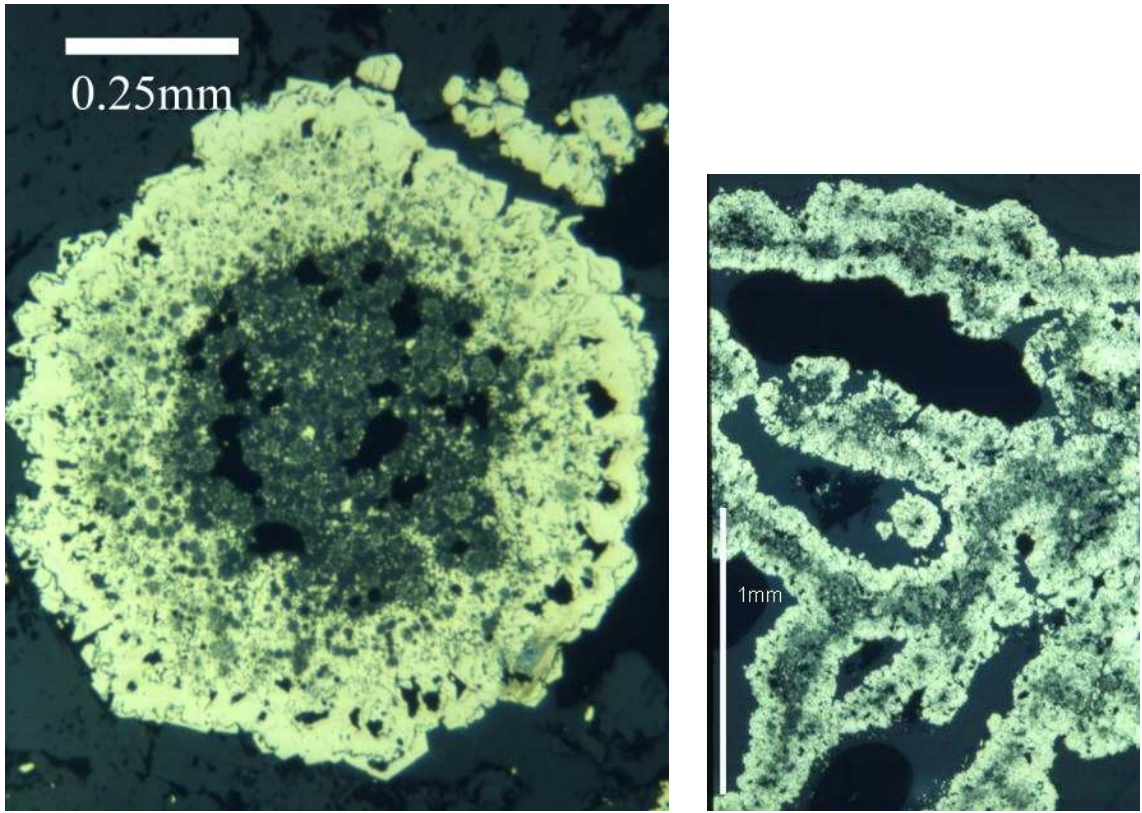

          (a) Small vent           (b) Sheaves

Figure 1. Framboids in chimneys: (a) Small pyrite vent structure: Reflected ore microscopy of transverse section shows a central area of empty black spaces plus (grey) fine framboidal pyrite, and a fine euhedral authigenic rim surrounded by baryte, with minor pyrite; (b) Sheaf system, formed from coalescing rods of anastamosing microcrystalline pyrite. Black areas are empty spaces; central regions are framboidal pyrite with an exterior of crystalline pyrite. (Labelled pictures given by Dr. Adrian Boyce are reproduced with his kind permission; Source: Boyce et al. 1983; Boyce 1990: Exhalation, sedimentation and sulphur isotope geochemistry of the Silvermines Zn + Pb + Ba deposits, County Tipperary, Ireland; Boyce, Unpublished Ph.D. thesis, University of Strathclyde, Glasgow).

     Since ordering and other dynamics in a system depend on its boundary conditions, it is reasonable to suppose that any molecular organization emerging in the mound as a result of evolving complexity of abiogenics and symbiotic interactions within them would be driven by the non-equilibrium sources therein. Then in the absence of gradients, upon swimming away from the confines of the mound, a suitable boundary condition is needed to sustain their organization, as in today's living systems sustained on homogeneous sources of energy. Hence for clues we look at present-day bacteria, using magnetosomes as compasses (Pósfai et al 1998; 2001; 2006) that help to orient them to the geo-magnetic field, as they navigate through waters. Interestingly, they are composed of ordered magnetite crystals, as well greigite ones, albeit less frequently (Reitner et al 2005; Simmons et al 2006). Their ancient origins can be appreciated from the conjecture

that "Magnetoreception may well have been among the first sensory systems to evolve" based on the universal presence of single-domain crystals of magnetite across many species and groups of organisms, ranging from bacteria through higher vertebrates that exploit the geo-magnetic field for orientation, navigation, and homing (Kirschvink et al 2001; Kirschvink and Hagadorn 2000). In addition, different magnetite-based models of biological magnetic-torque transducers have been proposed as a basis for sensing and transducing magnetic-field changes via direct/indirect coupling to mechanosensitive ion-channels (Winklhofer and Kirschvink 2010). Furthermore, to quote from Kopp and Kirschvink (2008), "The magnetosome battery hypothesis suggests the possibility that the magnetosome arose first as an iron sulfide-based energy storage mechanism that was later exapted for magnetotaxis and still later adapted for the use of magnetite," all of which have promising implications for magnetic mechanisms towards expanding the scope of the mound scenario.

In fact, Russell and coworkers (1990) have noted the size similarities of the magnetosome crystals to that of pyrite crystallites (~ 100nm in diameter) comprising the interior of framboids that appeared to have grown inorganically from spherical shells of iron-sulphide gel; their earlier stages would have comprised reduced forms of framboidal iron sulphide, just as in iron sulphide bearing bacteria. Revisiting the early enzyme FeS clusters then reveals that spin polarization and spin coupling are key characteristics of the sulphur bridged complexes, embedded within these ancient bio-constructs (Noodleman and Case 1992). The spins on metal sites in di- to polynuclear iron sulphur clusters are coupled via what is called Heisenberg exchange coupling, which typically favors anti-ferromagnetic alignment of neighbour-spins. This in combination with valence delocalization within the cluster helps bring about coupling of different degrees of freedom and add to the complexity of the system (Noodleman and Case 1992). Ligand binding can thus affect not only the electronic distribution, but also the pattern of spin-alignment, and hence the net total spin state (Noodleman et al 2002). The complex web of interactions-- orbital interactions, electron delocalization and spin coupling-- in the iron-sulphur clusters (Noodleman and Case 1995) shows the likelihood of a similar profile for nano-particles of mineral greigite, which is magnetic (a great candidate for complexity!). Could this feature have helped in the emergence of environment-stable assemblies capable of migration, and which could have 'chaperoned' proto-metabolic cycles? Could it address the feature of control embedded in the coherent long range order of crystals (Cairns-Smith 2008), and yet retain the benefits of a soft colloidal gel organization for heterogeneous catalysis that has distinct advantages over bulk phase homogeneous solution chemistry (Kopelman 1989; Russell et al 1994)?

To that end, we consider the formation of stable micro-meter sized greigite framboids with a magnetic basis of assembly, which has been modeled using an interplay of negatively charged repulsive and magnetically attractive forces (the dominant term in net energy of interaction), where a size > 100nm would orient crystals to the weak geo-magnetic field ~ 70 microTesla (Wilkin and Barnes 1997). Additionally, in an investigation of biologically induced mineralization, scale-free framboidal greigites have been obtained from Black Sea sediments that are ordered clusters of octahedral crystals comprising $Fe_3S_4$-spinels (Their diameters are mostly ~ 2.1, 4.2, 6.3 or 8.4 micrometer, with the two intermediate ones predominating. Nested structures, building up from the smallest one, lead to the higher sized clusters) (Preisinger and Aslanian 2004).

Next, we look at two aspects of this multi-faceted phenomenon: (i) important role of thermal perturbations, such as in randomly orienting magnetic moments (appearance of symmetry) associated with super-paramagnetic particles; and (ii) the role of single-domain particles, which besides their magnetization, are also characterized by their anisotropy energy. In this regard, the classical Stoner-Wohlfarth (1948) model provides insights into how an applied field influences hysteretic rotation of the magnetization over the magnetic-anisotropy barrier, and Neel (1949) studied how thermal agitation can help magnetization overcome the energy barrier. And Hoffman's (1992) observations of natural greigite crystals suggest that the single-domain to multi-domain transition occurs at a particle size of about 1 micrometer, below which the collision frequencies of these (assumed as) independently moving colloidal particles are controlled mainly by Brownian motion. Greigite particles are expected to be super-paramagnetic ~ 30-50nm size (Hoffman 1992), and can be used to define the lower limit of single-domain stability, below which the moment associated with the ferrimagnetic particle fluctuates as a result of thermal perturbations. As a result not only the magnetization of such a fixed particle can rotate, but also conversely the particle can be allowed to rotate even as its magnetic moment remains aligned to an external field (Chikazumi 1964, Wilkin and Barnes 1997). Hence this Néel superparamagnetic behaviour becomes relevant only if the particle is rigidly fixed in a matrix, and it seems reasonable to extend this possibility to greigite of this size embedded in the inorganic membranes (chemical gel), or even rigidifying framboidal networks (Larter et al 1981). On the other hand, the magnetic remanence associated with a newly forming aqueous suspension (Russell and Hall 1997) of super-paramagnetic and single-domain greigite particles--free-to-rotate in a viscous medium-- would have in principle access to both Neel and Brownian modes of relaxation. The latter characterizing the viscous rotation is proportional to the particle/assembly volume, in contrast to the former which is an exponential function of the volume. Therefore the Brownian mode for return to equilibrium becomes the dominant process for large single-domain particles suspended in liquid medium. The essential point is that magnetic dipolar interactions being weak, they provide a *reversible* interaction mechanism, important for feedback effects as will be taken up below. For now, being interested in only the general features expected in the mound with magnetic elements, we ignore complex situations (e.g. as the dipolar interaction between two neighbouring particles increases with decrease in intercrystal distance, the particle's aggregation-state should have an effect on the Neel relaxation, due to the dipolar intercrystal coupling aspect of the anisotropy (Laurent et al 2008)).

Now, the observations of scale-free greigite framboids shows the possibility of formation of nano-scale aggregates (c.f. Sawlowicz 1993; 2000) if the external field strength could be increased roughly to the milli-Tesla level, using the Wilkin-Barnes (1997) model (Mitra-Delmotte and Mitra 2010a). Indeed, magnetic nano-particle aggregation in response to an external field is a familiar feature of tunable magnetorheological fluids, resulting in a reversible and rapid increase in viscosity that is tunable by varying the field (Silva et al 1996). The synthetic dispersions of magnetic nano-particles in solvents are stabilized against agglomeration by surface-coatings (see Odenbach 2004), but structures that are formed in the presence of moderate fields, break up when the field is switched off. Interestingly, these structures can be maintained upon removal of field if magnetic dipolar interactions are additionally reinforced by weak chemical interactions such as Van der Waals or electrostatic bonds (Martínez-Pedrero 2008), and which are likely to have been present due to the organics stabilizing greigite particles (an otherwise metastable mineral phase, forming as an intermediate in FeS transformations (Rickard et al 2001; Russell and Hall 2006)). Weak short range interactions are ubiquitous in biology, enabling reversible transformations. Local pockets of disorder in magnetic dipolar ordering could have been introduced using weak chemical bonds; and as system states would get entangled in the history of the aggregation process, this may have created *uncertainty* amidst *order*. Further, Wilkin and Barnes (1997) argue that surface-minimization causes spherical rather than chain-like structures; note also that spherical aggregates form in weak fields but elongated ones in stronger magnetic fields. Roughly guided by these observations and that of scale-free framboids by Sawlowicz (1993), it seems logical that moderate fields would allow spherical aggregates at the nano-scale, but stronger ones would lead to higher aspect ratios (see Wilkin and Barnes 1997 for details). Furthermore, fractal greigite framboids (Preisinger and Aslanian 2004) independently show that this mineral can give access to an aperiodic, nested organization.

Intuitively, a magnetic field (c.f. field-induced phase transition, Haken 2004) seems the perfect homogeneous ordering energy source relevant for life, especially for a fitting explanation for symmetry-breaking, an aspect living systems have taken to new dimensions (Anderson 1972). In particular, its orienting mechanism, which effectively lowers the entropy of a dispersion of particles with thermally fluctuating moments and enables their dipolar ordering, closely resembles the ordering role of energy in living systems (c.f. "drinking orderliness" Schrödinger 1944), with a polar, asymmetric organization (Harold 2005) and complementary recognition-based communication (Mitra-Delmotte and Mitra 2010b). Now, the fact that there was no geo-magnetic field ~4.1-4.2Gya (Hazen et al 2008), by when life is already seen to have been initiated, i.e. ~4.2-4,3 Gya (Russell and Hall 1997; Russell and Hall 2006), leave alone the requirement of the order of tens of milli-Tesla for stabilizing nano-scale aggregates (see above), might seem discouraging. Nevertheless, a magnetic field could be imagined in a local manner, such as assuming the presence of extra-terrestrially magnetized rocks (Mitra-Delmotte and Mitra 2010a). A moderate field could then enable the aqueous suspension of super-paramagnetic greigite particles, forming in the mound (Russell and Hall 1997) to overcome thermal fluctuations and align to the field. The consequent assembly into nano-framboid-like aggregates is speculated on the grounds of Sawlowicz's (2000) theory based on empirical observations of fractal framboids; the observations of a nested hierarchy of greigite framboids (Preisinger and Aslanian 2004), showing that greigite can provide access to such scale-free order; resemblance of this scenario to field-induced aggregates in synthetic magnetic-nanoparticle dispersions (see above); and the fact that framboids have been observed in chimneys (see Figure 1; Boyce et al. 1983; Boyce 1990; Larter et al. 1981).

It is also relevant to note the similarity between the passage of a magnetosome moving in response to the extremely-low-frequency geo-magnetic field (Kirschvink et al 1992) (although by sensing changes in the inclination magnitude increasing from the equator to the poles) and the *scaled-down scenario* of ligand-bound super-paramagnetic greigite nano-particles traversing a field-induced aggregate in response to (gentle changes in flux lines due to) an external moderate H-field (Mitra-Delmotte and Mitra 2010a). Next, for internal dynamics in the latter, binding to organic ligands would be expected to increase the potential barrier for the dipolar interaction. Thus greater diffusive exploration of the organic bound particles (as in bio-molecular motors) contrasts with the entrapment of the unshielded ones into an expanding network of dipolar interactions (as in growth phenomena). A consequence of this asymmetric dynamics is the accumulation of chiral organics resulting from transfer reactions between ligands bound to the migrating particles; these in turn would have been gradually incorporated into the network en route to an asymmetric organic one. These field-induced assemblies do not depend on the mound-boundary-conditions (Russell and Arndt 2005; Russell et al 2005) for their ordering (magnetization of particle dependent on field strength), so in principle they offer a substitute for stable membrane organization even in the absence of these boundary conditions (see above). And even with such variations, such as in gradient-rich niches away from the mound that would be required for driving metabolism, as well as molecular dynamics/organization exploiting the complexity of evolved abiogenics (e.g. thermophoretic effects (Braun and Libchaber 2004)), the 'chaperoning' magnetic jacket hosting these processes could use changes in susceptibility in response to such intensive parameters (temperature, chemical potential, etc) towards generation of different energy transduction mechanisms. These magnetic assemblies provide a logical scenario for bringing together separately-evolved mechanisms harnessing a variety of fluxes, since further colloidal/mineral precipitates could envelop the mound.

Analogous to crystal lattice vibrations (Cairns-Smith 2008), it could be reasonable to suppose that spin changes in particles (due to thermal fluctuations; catalyzed redox reactions; ligand-effects, etc) would affect their magnetization and consequently their dipolar network. The reversible nature of latter offers a mechanism for feedback, e.g. for the kinetics of a chemical reaction occurring on the network, leading to favourable/unfavourable products to be 'judged' in terms of their effect on the stability of the system (c.f. "crystal genes" (Cairns-Smith 2008)). A magnetic-scaffold also gives its supporting local magnetic-field the role of a selective environment for sustaining its gradual asymmetric organic replacements. Consider, e.g., the "independence assumption" (Schneider 1991): it works just as well for the collapse of a gas-like compressed phase of a dissipative assembly of non-interacting particles in the presence of an H-field (Li et al 2007), as it does for the important components of a sophisticated molecular machine modeled as an ideal gas in a suitably defined higher dimensional space (Schneider 1991). This is *not* to compare the information-storage-complexity of these vastly different systems but only to show how much a big principle like the "independence assumption" can go towards unifying their common features. This shows that even if the physical substrate embodying the process is different, the similarity of operations in the processes embodied by different physical substrates does make it logical to consider that a field could have used this as a selection-mechanism to differentiate and choose the organic candidates fit to replace the components of a mineral-scaffold (c.f. Cairns-Smith 1982; 1985; 1992), in order to maintain the continuity of functions being carried out by inorganic minerals. We wonder if their entry could not be reasoned via multi-ferroic effects (Khomskii 2006), as these have been found to be relevant for minerals (Hemberger 2006), and in view of the known correspondences between solid state physics processes and biological ones (Cope 1975). This is interesting in view of the analogy of electro-rheological to magneto-rheological fluids where dielectric particles replace paramagnetic ones, although in biology today, similar effects seem to be intricately entangled. (For instance, in addition to coherent electric dipole ordering alignment of actin monomers prior to ATP-activation, Hatori et al (2001) report the coherent alignment of magnetic dipoles induced along the filament, by the flow of protons released from ATP molecules during their hydrolysis).

We conclude this discussion, asking if magnetism could have provided a *coherent* basis for scaling-up the soft-matter replacements of these magnetic networks above the nano-scale size, by analogy to magnetic framboids, so that the weak newly-formed geo-magnetic field, by 3.4 Gya (Tarduno et al 2010) could have helped sustain colloidal assemblies outside the local confines of the rock-field. Indeed, this leads us to Frohlich's (1968) suggestion that order in biological systems may be of the same nature as that in coherent systems like lasers. According to this, upon receiving a continuous supply of metabolic energy (above a threshold), a non-linear interaction between two or more types of d.o.f.s of the ordered system acting like coupled molecular oscillators in a heat bath (given a mechanism for stabilizing the properties of the excited state) could lead to the emergence of a strongly excited, single mode, similar to Bose-Einstein condensation. Although this was in the context of collective dipolar modes associated with periodic bio-molecular structures, it is intriguing to consider a similar extrapolated-back-scenario for the evolving H-field-supported system (see above) in search of the feature of coherent-order, implicit in "crystal genes". Recent work by one of us (Mitra 2010) shows that while decoherence is usually positive for most two-way couplings with the environment, a three − way interaction involving the system, bath and H-field all together, leads to a reversal of sign of this quantity, which indicates its potential to provide support/sustain coherent phenomena.

**Acknowledgements**: We are indebted to Prof. MJ Russell for inspiration, apart from providing data plus key references, and a critical reading of the manuscript. We are grateful to Dr. Adrian Boyce for generously providing his labeled framboid pictures; and to Dr. Jean-Jacques Delmotte for full infrastructural support.